\documentclass{aa}  
\usepackage{amsmath}
\usepackage{graphicx}
\usepackage{graphics}
\usepackage{subfigure}
\usepackage{txfonts}
\usepackage{natbib}
\bibpunct{(}{)}{;}{a}{}{,}

\def\ltsima{$\buildrel<\over\sim$}
\def\lsim{\lower.5ex\hbox{\ltsima}~}
\def\gtsima{$\buildrel>\over\sim$}
\def\gsim{\lower.5ex\hbox{\gtsima}~}

\def\teff{\ifmmode T_{\rm eff} \else $T_{\mathrm{eff}}$\fi}

\def\lya{Ly$\alpha$} 
\def\mclya{{\em McLya}}

\def\ebv{\ifmmode E_{B-V} \else $E_{B-V}$\fi}

\def\fesc{\ifmmode f_{\rm esc} \else $f_{\rm esc}$\fi}

\def\cm2{cm$^{-2}$}
\def\kms{km~s$^{-1}$}

\def\ewlya{$EW_{\mathrm{Ly}\alpha}$}

\def\hi{H{\sc i}}

\def\nh{\ifmmode N_{\mathrm{HI}}\else $N_{\mathrm{HI}}$\fi}
\def\nhi{\ifmmode N_{\mathrm{HI}}\else $N_{\mathrm{HI}}$\fi}

\def\vexp{\ifmmode {\rm v}_{\rm exp} \else v$_{\rm exp}$\fi}
\def\taua{\ifmmode \tau_{a}\else $\tau_{a}$\fi}

\begin{document}
  \title{Grid of \lya\ radiation transfer models for the interpretation of distant galaxies}
  \subtitle{}
  \author{Daniel Schaerer\inst{1,2}, Matthew Hayes\inst{1,2}, Anne Verhamme\inst{3}, Romain Teyssier\inst{4,5}}
  \institute{
Observatoire de Gen\`eve, Universit\'e de Gen\`eve, 51 Ch. des Maillettes, 1290 Versoix, Switzerland
         \and
Laboratoire d'Astrophysique de Toulouse-Tarbes, 
Universit\'e de Toulouse, CNRS,
14 Avenue E. Belin,
31400 Toulouse, France
\and Universit\'e de Lyon, Lyon, 69003, France; Universit\'e Lyon 1, Observatoire de Lyon, 9 avenue Charles Andr\'e, Saint-Genis Laval, 69230, France ; CNRS, UMR 5574, Centre de Recherche Astrophysique de Lyon ; Ecole Normale Sup\'erieure de Lyon, Lyon, 69007, France
\and Universit\"at Z\"urich, Institute f\"ur Theoretische Physik, 
Winterthurerstrasse 190, 8057 Z\"urich, Switzerland
\and Laboratoire AIM, CEA/DSM - CNRS - Universit\'e Paris Diderot, IRFU/SAp, 
91191 Gif sur Yvette, France
 }

\authorrunning{}
\titlerunning{\lya\ radiation transfer models}

\date{Accepted for publication}

\abstract{\lya\ is a key diagnostic for numerous observations of distant star-forming galaxies.
It's interpretation requires, however, detailed radiation transfer models.}
{
We provide an extensive grid of 3D radiation transfer models simulating the \lya\ and 
UV continuum radiation transfer in the interstellar medium of star-forming galaxies.}
{
We have improved our Monte Carlo \mclya\ code, and have used it to 
compute a grid of 6240 radiation transfer models for homogeneous spherical 
shells containing \hi\ and dust surrounding a central source. The simulations cover a 
wide range of parameter space. We present the detailed predictions from our models including 
in particular the \lya\ escape fraction \fesc, the continuum attenuation, and detailed \lya\ line 
profiles.}
{
The \lya\ escape fraction is shown to depend strongly on dust content, but also on other 
parameters (\hi\ column density and radial velocity). The predicted line profiles show a great
diversity of morphologies ranging from broad absorption lines to emission lines
with complex features. 
The results from our simulations are distributed in electronic format.}
{
Our models should be of use for the interpretation of observations from distant 
galaxies, for other simulations, and should also serve as an important base for 
comparison for future, more refined, radiation transfer models.}

  \maketitle

\section{Introduction}

Stimulated by numerous observations of galaxies and the intergalactic medium
(IGM) in the high redshift Universe, and made possible by significant growth of
computational power, several groups have recently developed numerical codes 
treating the transfer of \lya\ radiation 
\citep[e.g.]{zheng02,ahn03,dijkstra06a,verhamme06,tasitsiomi06,laursen07,barnes09}.
Most of the codes use the Monte Carlo (MC) method to sample the radiation
transfer. The main advantages of this technique is its ease in implementation,
treatment of arbitrary geometries, and inclusion of different physical
scattering and absorption processes.
The codes have been applied to a variety of astrophysical problems,
including e.g.\ cosmic reionisation, \lya\ fluorescence in the IGM,
predictions of \lya\ emission from distant galaxies,
and the interpretation of observed \lya\ line profiles from starburst galaxies
\citep{cantalupo05,cantalupo08,semelin07,baek09,pierleoni09,barnes10,kollmeier10,
laursen11,zheng11}.

Our group has recently developed a general 3D radiation transfer code
including \lya\ and UV continuum radiation including scattering and absorption
by neutral hydrogen and dust \citep{verhamme06}.
So far, our code has mostly been applied to understand the diversity
of observed \lya\ line profiles and to quantitatively exploit the
\lya\ profile in \lya\ emitters (LAE) and Lyman Break Galaxies (LBG)
\citep[see][]{schaerer08,verhamme08,dessauges09,vanzella10} and for the modeling of local
starburst galaxies \citep[cf.][]{atek09a}. 
In particular, we have shown that the observed diversity of line profiles 
reaching from broad absorption in some LBGs to asymmetric emission
in LAEs, can be understood by variations of the \hi\ column density
and dust content in spherically expanding shells. 
Furthermore our models also place interesting constraints on the intrinsic
\lya\ emission properties, hence on age and star-formation history.
For example, for LBGs we found no need for particular ages, and we showed that 
their intrinsic \lya\ equivalent widths are compatible with constant star-formation
over several hundred Myr, as also indicated by their broad-band SEDs.
Last, but not least, our simulations have clarified the relation
between observed velocity shifts of \lya\ and ISM expansion velocities 
\citep[see][]{schaerer08,verhamme08}. 

These studies have been based on a relatively small number of simulations
and tailored models to fit observations of less than twenty galaxies,
albeit of different types. To explore a larger range of the parameter
space, and to do this in a systematic way, allowing us also to examine
possible degeneracies in \lya\ line profile fits, we have computed
a large grid of radiation transfer models for homogeneous, spherically
expanding shells. 
Our simulations cover a wide range of the 4-dimensional parameter space
given by the expansion velocity, \hi\ column density, velocity dispersion 
($b$), and dust optical depth. From this, spectra can be predicted
for any arbitrary input spectrum, containing e.g.\ a UV continuum
plus intrinsic \lya\ emission from the central source.

To make these simulations feasible, we have parallelised our Monte Carlo
code, named \mclya. Furthermore we have added some new physics in the code, such as
scattering by Deuterium and the recoil effect, and we have made other 
improvements (in particular on scattering by dust).
Our standard code works on cartesian grids; a version using nested-grids, 
provided by AMR (Adaptive Mesh Refinement) codes has recently been 
developed by Verhamme et al. (in preparation).
%
In addition we have developed an automated tool to fit observed
\lya\ line profiles, making use of our grid of \mclya\ simulations.

The description of the grid of UV and \lya\ radiation transfer simulations
and some results derived from it, are the main objective of this paper.
At the same time the model grid described here will be made available 
electronically.

Predictions from our model grid have already been used in several
papers. For example, \citet{atek09b} compared predicted \lya\ escape fraction from
our grid to measurements of this quantity in $z\sim$ 0.2--0.4 galaxies, and
\citet{hayes10b,hayes10c} have compared the same quantity to measurements
at redshift $z \sim 2$ and higher.
In \citet{dessauges09,vanzella10}, we have used our fitting tool to
analyse two individual LBGs at redshift $z=2.8$ and $z=5.8$ respectively.
The predictions from our model grid have recently been included in the semi-analytical
galaxy models of the Lyon group \citep{garel11}. Our predictions are also being 
used to prepare observations with future instruments, such as MUSE for the VLT
(Garel et al., in preparation). 

Our paper is structured as follows.
In Sect.\ \ref{s_mclya} we describe the main improvements implemented in our code.
In Sect.\ \ref{s_grid} we summarise the input parameters and other issues
for the computation of our models grid.
Predictions for the \lya\ escape fraction and the strength of \lya\ absorption 
are shown in Sect.\ \ref{s_esc}. 
In Sect.\ \ref{s_profile} we illustrate some line profiles predicted from our
model grid.
In Sect.\ \ref{s_conclude} we summarise our main conclusions.

\section{Update of the MCLya code}
\label{s_mclya}

We have developed an improved version of the Monte Carlo radiation transfer code \mclya\ of
\citet{verhamme06} including the detailed physics of \lya\ line and UV continuum transfer, 
dust scattering, and dust absorption for arbitrary 3D geometries and velocity fields.

The following improvements have been included:

\begin{itemize}
\item Angular redistribution functions taking quantum mechanical results for \lya\ into account.
More precisely we distinguish two different phase functions for scattering in the core
and the wing, following \citet{dijkstra08}. For the distinction between these two regimes
we also use their value of $x_{\rm crit} = 0.2$, where $x$ is the frequency shift from
line center expressed in Doppler units.
In the wing, the phase function is identical to the one of dipolar (Rayleigh) scattering,
adopted previously in our code for the computations of \citet{verhamme08}.
Although the differences in the phase functions are important for a proper
treatment of polarised radiation \citep[cf.][]{dijkstra08}, we have not noticed 
changes in our predictions.

\item Frequency changes of \lya\ photons due to the recoil effect \citep[e.g.][]{zheng02}.
Before, scattering was considered coherent in the atom's frame.
As well known, the recoil effect can lead to differences for very low temperatures,
as shown e.g.\ by \citet{zheng02,tasitsiomi06}.

\item The presence of Deuterium.
Following the suggestion of \citet{dijkstra06a}, Deuterium has been included, assuming a canonical 
abundance of $D/H = 3 \times 10^{-5}$. It's effect may be visible for cases with 
a static ISM in the blue part of the line \citep[cf.\ Fig.\ 3 of][]{dijkstra06a}. 
For expanding geometries, such as the ones discussed below, it's effect becomes, however,
very small or invisible.

\item Anisotropic dust scattering using the Henyey-Greenstein phase function.
We adopt the following values for the dust albedo $Q_a=0.46$ and the scattering phase
function asymmetry $g= \langle\cos \theta\rangle =0.77$ at \lya\ following \citet{witt00}.
In our previous computations we adopted $Q_a=0.5$ and $g=0$ (isotropic dust scattering).
Among the changes made, this improvement has the largest impact on the results
since it leads to a more pronounced fore-aft symmetry for dust scattering, which
in turn affects the escape fraction of photons both in the UV continuum and in the 
\lya\ line.
For the spherical shell geometry considered below, this leads in general to somewhat
higher escape fractions, i.e.\ less efficient dust attenuation.

\item 
Finally, the code has been parallelised using OpenMPI, for efficient use on supercomputers.
Given the use of the Monte Carlo method, parallelisation is basically trivial,
and near-perfect scaling properties are achieved. Typically, we have run our simulations on 
100-300 CPU cores in parallel.
\end{itemize}

The remaining input physics is described in \citet{verhamme06}.
We do not rely on approximations to accelerate the \lya\ transfer.
Although we have experimented with some methods \citep[e.g.\ those in][]{verhamme06,laursen09},
we have noted some differences in the resulting line profiles. 
Therefore, to avoid possible inaccuracies in our calculations exploring a large
parameter space, we prefer to use a correct, ``brute force'' approach.

\section{Grid of MCLya models}
\label{s_grid}

\begin{table*}
\caption{Input parameter values adopted for the grid of radiation transfer models
of spherically expanding shells. All combinations of parameters have
been computed, corresponding to 6240 models in total.
The units are the following: \vexp\ and $b$ in \kms, \nhi\ in \cm2. \taua\ is dimensionless.
By symmetry of the problem, the models also describe spherical infall, in which case 
the infall velocity is $v= - \vexp$.}
\label{t_grid}
\begin{tabular}[htb]{llrrrrrll}
Parameter & Values & $n$
\smallskip \\
\hline 
\vexp        & 0.,20., 50., 100., 150., 200., 250., 300., 400., 500., 600., 700. & 12 \\
$\log(\nhi)$ & 16., 18., 18.5, 19.,19.3,19.6,19.9,20.2,20.5,20.8,21.1,21.4,21.7 & 13 \\
$b$          & 10., 20., 40., 80., 160. & 5 \\
$\taua$      & 0., 0.2, 0.5, 1., 1.5, 2., 3., 4. & 8 \\
\hline
\end{tabular}
\end{table*}

For simplicity, and to provide a basis for other future studies,
all simulations carried out subsequently assume a homogeneous and co-spatial 
distribution of neutral hydrogen and dust with a constant density and temperature.
The effects of clumpy media are examined in an upcoming publication (Duval
et al., in preparation).
The input parameters of the code are the \hi\ geometry and velocity field,
the microscopic \hi\ velocity distribution,
the spatial location and distribution of the UV continuum and line emission source(s),
and the dust-to-gas ratio.

For the current model grid we consider the following geometry:
spherically symmetric shells with a central source UV source.
This case is described by 4 parameters:
{\em (i)}   the radial expansion velocity of the shell, \vexp, 
{\em (ii)}  the \hi\ column density towards the source, \nhi, 
{\em (iii)} the microscopic \hi\ velocity distribution described by the Doppler parameter $b$, and
{\em (iv)} the dust absorption optical depth \taua, which expresses the dust-to-gas ratio.
As discussed by \citet{verhamme06} \taua\ is related to the usual color excess
$E(B-V)$ by  $E(B-V) \approx (0.06...0.11) \taua$. 
For the \citet{calzetti00} law one has $E(B-V) \approx 0.1$ \taua.
In short, the present model grid is described by 4 input parameters \vexp, \nhi, 
$b$, and \taua. 
By symmetry of the radiation transfer problem \citep[cf.][]{neufeld90,verhamme06,dijkstra06a}
our grid calculations also apply to the case of spherical infall. In other
words models with a radial expansion velocity \vexp\ also describe the 
case of spherical infall with $v = - \vexp$.

Thanks to the parallelisation of the code, the computation of a large
grid, covering a wide 4-dimensional parameter space, has been
possible.  The values adopted for the input parameters are listed in
Table \ref{t_grid}.  Simulations have been run for all combinations,
yielding in total 6240 models. 

For each parameter set, a full Monte Carlo simulation is run with 1000 photons
per input frequency bin. The radiation transfer calculations cover a
sufficiently broad spectral range (here typically from $-6000$ to
$+6000$ \kms, in bins of 20 \kms\ for the input) to reach the
continuum for most simulations (except for some of the highest column 
density simulations). 
As described in \citet{verhamme06}, our MC simulations are computed for a flat input
spectrum, keeping track of the necessary information to recompute 
{\em a posteriori} simulations for arbitrary input spectra.
The total computing time required for the entire grid amounted to $\approx$ 25 
CPU year.

In practice, the following quantities are stored for each emergent photon:
input and output frequency, position and direction of escape.
From this we can in particular compute the predicted emergent spectrum (integrated 
or spatially resolved spectra, spectral maps) and the escape fraction
for all frequency bins. The detailed model results (approx.\ 60 GB)
allowing one to compute emergent spectra are available on request from the 
first author\footnote{See {\tt http://obswww.unige.ch/sfr} or contact DS.}. 
Derived quantities (the \lya\ escape fraction
computed for different FWHM, the continuum escape fraction, and the
equivalent width of the \lya\ absorption assuming a flat continuum, cf.\ below)
for all models are provided in Table \ref{t_fesc}.

\begin{table*}
\caption{Derived quantities \fesc, $\fesc^{\rm cont}$, and \lya\ equivalent width
for a constant input spectrum (flat continuum) (cols.\ 6-8) computed for all 6240 models 
(described by the parameters given in cols.\ 1-4) and for different
FWHM values of the input \lya\ line (col 5).
Note that the derived quantities are identical for cases of spherical infall
with the same absolute radial velocity ($|\vexp|$).
The full table is available only in the electronic version or on request
from the author.}
\label{t_fesc}
\begin{tabular}[htb]{llllllllll}
$b$ & $\taua$ & $\log \nh$ & \vexp\ & FWHM(\lya) & \fesc\ & $\fesc^{\rm cont}$ & EW \\ 
\protect[ \kms ] &  &  [\cm2 ] & [\kms ] & [\kms ] &  &  & [\AA ] \\
\smallskip \\
\hline 
$\ldots$\\
10. &     0.2 &    16.0 &     0. &    50. &  0.41473 &  0.84103 &   -0.13742\\
10. &     0.2 &    16.0 &     0. &   100. &  0.57985 &  0.84103 &   -0.13742\\
10. &     0.2 &    16.0 &     0. &   150. &  0.65799 &  0.84103 &   -0.13742\\
10. &     0.2 &    16.0 &     0. &   200. &  0.70028 &  0.84103 &   -0.13742\\
$\ldots$\\
\hline
\end{tabular}
\end{table*}
\begin{figure}[tb]
\centering
\includegraphics[width=8.8cm]{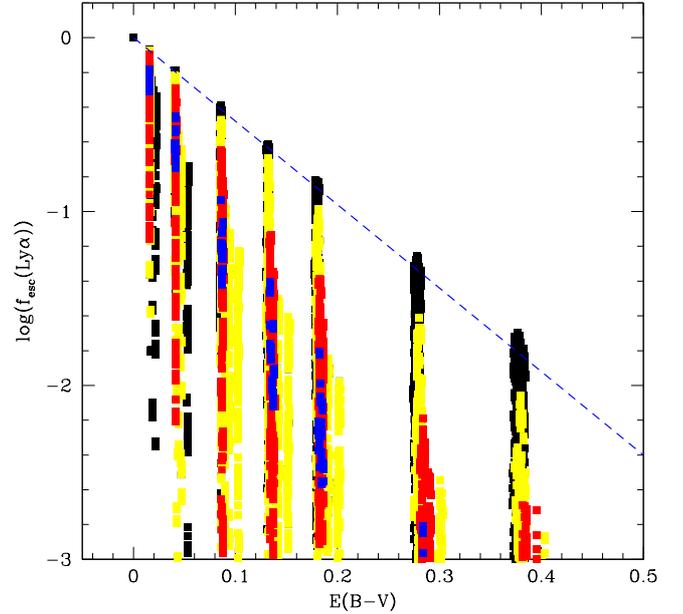}
\caption{Predicted \lya\ escape fractions from our 6240 models as a function 
of \ebv.
Yellow (red) symbols show the simulations with dust-to-gas 
ratios within a factor 10 (2) of the Galactic value.
Among the models with dust-to-gas ratios within a factor 2 of the Galactic value,
those with \vexp=200 \kms\ are shown by blue symbols.
Black symbols show all the remaining models. The blue dashed line shows
an attenuation with $k_\lambda=12.$ corresponding to the \citet{calzetti00} law.
Values above this line are due to numerical noise.
}
\label{fig_fesc_ebv}
\end{figure}

\begin{figure}[tb]
\centering
\includegraphics[width=8.8cm]{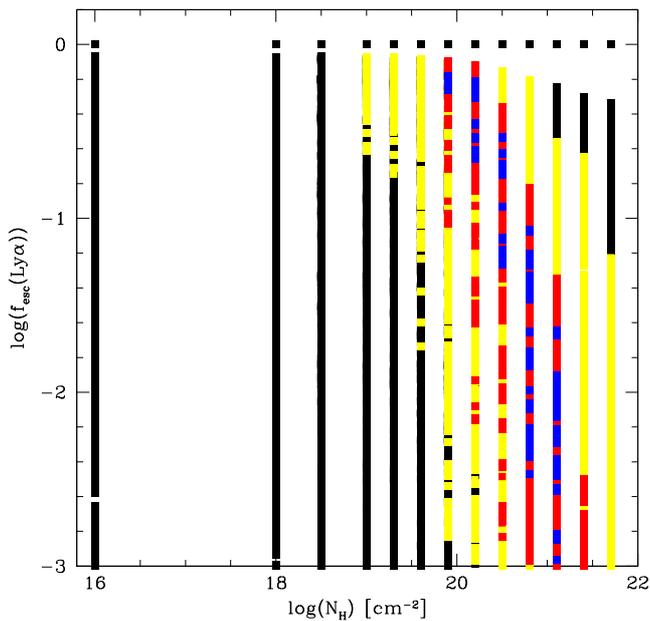}
\caption{Predicted \lya\ escape fractions from our 6240 models as a function 
of \nh. Colour codes for the symbols as in Fig.\ \protect\ref{fig_fesc_ebv}.}
\label{fig_fesc_nh}
\end{figure}
\section{Predictions for the \lya\ escape fraction and related quantities}
\label{s_esc}

The total \lya\ escape fraction, \fesc, is defined as the ratio between the
number of \lya\ line photons emitted, $N$, by the UV source and the number of
these photons, $N_{\rm esc}$, emerging from the simulation box. 
The monochromatic escape fraction is given by
$\fesc(\lambda) = N_{\rm esc}(\lambda)/N(\lambda)$, 
where $\lambda$ is the ``input'' wavelength, i.e.\ that of the photons
emitted from the source before being altered by radiation transfer.
The total  \lya\ escape fraction is
\begin{equation}
\fesc = \frac{\int N_{\rm esc}(\lambda) d\lambda}{\int N(\lambda)d\lambda},
\label{eq_fesc}
\end{equation}
where the integration is carried out over the entire line profile
of the ``input'' spectral line.
In principle \fesc\ therefore depends on the emission line profile
of the source. In practice, however, \fesc\ is close to the monochromatic
escape fraction at line center $\fesc(\lambda_0)$ for reasonable widths
of the emission line profile (typically \fesc\ varies by less than
0.2 dex for FWHM(\lya) between 50 and 200 \kms),
since the resulting \hi\ absorption line profile is broader than FWHM for most
column densities.
In any case, we compute \fesc\ assuming a Gaussian emission line profile
with varying FWHM(\lya) from 50 to 1000 \kms\ for various applications. 
These values are provided in Table \ref{t_fesc}.

\subsection{Dependence of \fesc\ on the physical parameters}
Since \lya\ photons can only be destroyed by absorption by dust particles,
\fesc\ mainly depends on the dust optical depth, described here by \taua.
The predicted \lya\ escape fraction for all models is shown as a function
of \ebv\ in Fig.\ \ref{fig_fesc_ebv} for lines with input FWHM $\le 200$ \kms.  
Here \ebv\ is derived from the escape fraction in the continuum near \lya, assuming $k_\lambda=12$ and
$R_V=4.05$ from the \citet{calzetti00} law.
Very low escape fractions are obtained in some models.
In such cases the line profiles are dominated by a broad absorption
(Voigt-like profile) and no distinctive trace of the intrinsic \lya\ emission line
is detectable. For this reason we limit our plots, somewhat arbitrarily, 
at $\fesc\ \approx 10^{-3}$. This limit also corresponds to our numerical
limit on the escape fraction per frequency bin.

The \lya\ escape fraction has an upper limit of 
$\fesc \la \exp(-\taua) \approx \exp(10 \times \ebv)$, corresponding to the
pure attenuation of the continuum. Below this value we see that \fesc\
varies by several orders of magnitudes, when variations of the remaining 
parameters (\vexp, \nh, $b$) over a large range are allowed.
In this case, multiple scattering effects of \lya\ on HI and on dust
increase the probability of subsequent dust absorption of \lya\ photons,
reducing therefore \fesc.
For example, for a given \taua (\ebv), increasing the HI column density, \nh,
leads to wide range of \fesc, allowing thus in particular low escape fractions.
Similarly, a nearly static ISM (low expansion velocities \vexp) leads
to lower \lya\ escape fractions, when all other parameters are the same.
In both cases the increase of \nh\ and the decrease
of \vexp\ increases the number of \lya\ (and dust) scatterings
due to radiation transfer effects, and hence the probability of
absorption by dust for same radial dust optical depth.
In particular we note that relatively low \lya\ escape fractions
($\fesc \la 0.1$ or less) can even be obtained in situations with very 
little dust (\ebv $\la$ 0.02), provided the ISM has low velocities
with respect to the UV source.

In reality not all combinations of parameters may be realised. For
example, the dust-to-gas ratio 
\begin{eqnarray}
[{\rm dust/gas}] & =&  \log(\ebv/\nh)- \log(\ebv/\nh)_{\rm Gal} \\ \nonumber
           & \approx& \log(0.1 \taua) -\log(\nh) + 21.76
\end{eqnarray}
may be restricted within some range from the Galactic value 
$\log(\nh/\ebv)_{\rm Gal}=\log(5.8 \times 10^{21})$ \cm2\ \citep{bohlin78}.
As clear from Table \ref{t_grid}, some of our models have extreme
dust-to-gas ratios both above and below unity.
In Figs.\ 1--4 we therefore distinguish with different colors 
the models which dust-to-gas ratios within a factor 10 (2) of the Galactic value,
i.e.\ with [dust/gas] $\in [-1,1]$ ([-0.3,0.3]), and the remaining models.
Among the models with dust-to-gas ratios close to the Galactic value
(within a factor 2; i.e.\ among the red points) we have further highlighted
in blue models with a typical outflow velocity of \vexp=200 \kms\
to illustrate how e.g.\ velocity affects the \lya\ escape fraction.
The remaining spread in \fesc\ still shows how the other parameters
($b$, \nh) affect the radiation transfer and hence the predicted \lya\ escape.

\begin{figure}[tb]
\centering
\includegraphics[width=8.8cm]{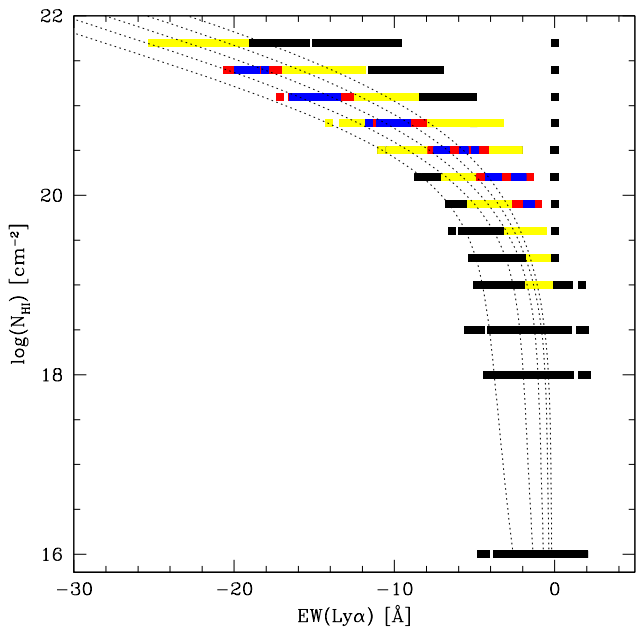}
\caption{Predicted \ewlya\ in absorption from our model grid as a function 
of \nh\ (black points) for a flat (constant) input spectrum,
simulating a pure UV continuum. 
Colour codes for the symbols as in Fig.\ \protect\ref{fig_fesc_ebv}.
Predictions at \ewlya$=0$ correspond to dust free models.
The dotted lines show \ewlya\ obtained from pure Voigt profiles
for $b=10$, 20, 40, 80, and 160 \kms\ (from right to left)
}
\label{fig_ewabs_nh}
\end{figure}

\begin{figure}[tb]
\centering
\includegraphics[width=8.8cm]{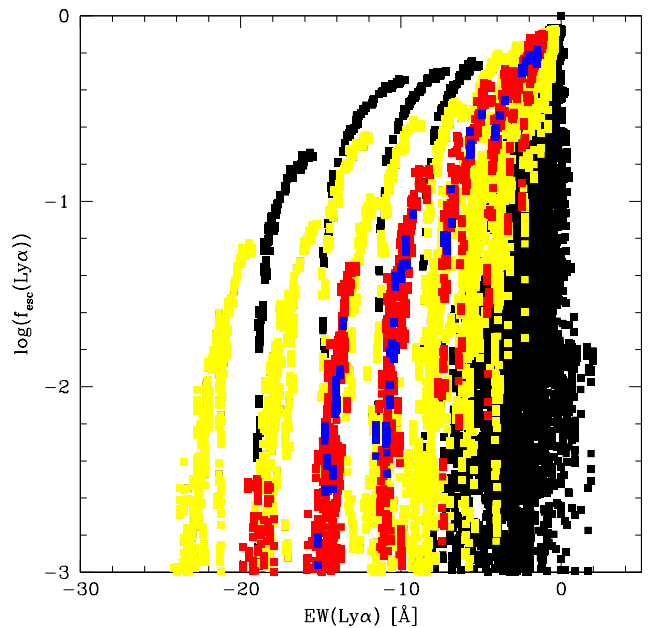}
\caption{Predicted \lya\ escape fractions from our 6240 models as a function 
of the absorption equivalent width expected for a flat (constant) UV
spectrum.
Colour codes for the symbols as in Fig.\ \protect\ref{fig_fesc_ebv}.
}
\label{fig_ewabs_fesc}
\end{figure}

\subsection{\lya\ absorption}
Radiation transfer effects not only regulate the transmission
of \lya\ line photons discussed above; they also affect the UV continuum 
photons, ``carving'' thus broad \lya\ absorption lines often
observed in spectra of distant objects.
Whereas scattering of photons out of the line-of-sight is
principally at the origin of absorption lines of the \lya\ forest,
true absorption by dust is the only processes capable of 
creating \lya\ absorption lines for the ``closed'' geometry
and for integrated spectra we consider here \citep[cf.][]{verhamme06}.

In Fig.\ \ref{fig_ewabs_nh} we show the predicted equivalent width
of \lya\ absorption ($EW<0$) from our model grid as a function of the 
\hi\ column density. Here \ewlya\ has been computed assuming a
constant (flat) UV spectrum and measuring $EW$ over the interval [-6000,6000] 
\kms\ covered by our models\footnote{For the highest column densities, 
the \lya\ line is broader than this. Measuring \ewlya\ over 
[-10000,10000] \kms\ instead, increases $EW$ by $\sim$ 10--20 \%.}.
Dust-free models show $EW=0$\footnote{Few models show \ewlya$>0$ but 
close to zero. This is due to numerical noise.}, whereas otherwise 
$|EW|$ increases broadly with \nh, as expected. 
For a given column density, the strength
of the \lya\ absorption increases with increasing dust-to-gas ratio 
(coded in color). It also depends on the other parameters affecting 
the radiation transfer, here mostly \vexp.

For comparison, $EW$ computed from Voigt profiles as function of \nh\
and $b$ (in the same fashion, i.e.\ over [-6000,6000] \kms) are shown 
by dotted lines in  Fig.\ \ref{fig_ewabs_nh}.
As discussed by \citet{verhamme06} and shown e.g.\ from
detailed line profile fitting by \citet{dessauges09}, radiation transfer models
for the geometry adopted here predict in some cases weaker \lya\ absorption
than expected for \nh\ deduced by simple Voigt profiles.
In other words, if applicable, our geometry could imply that 
\nh\ measured from pure Voigt profile fits are underestimated.
However, the precise amount of this difference depends on
the $b$ value adopted for the Voigt fits, on the dust-to-gas-ratio,
and on other parameters.

Overall we note that the strength of the \lya\ absorption
predicted by our model grid covers well the range of observed \ewlya\
values in Lyman Break Galaxies (LBGs), with equivalent widths
down to $\sim$ -20 to -30 \AA\ \citep[e.g.][]{shapley03}.

As expected, there is no strong correlation between the escape fraction
of \lya\ line photons and $EW_{abs}$ (see Fig.\ \ref{fig_ewabs_fesc}).
The former measures the transmission of the photons close to line 
center, the latter is an overall measure of the net transmission of the
UV flux. Therefore it is clear that cases with a strong absorption line
(\ewlya $\ll 0$) also show a low \lya\ escape fraction. However, 
for some models with low \hi\ column densities and high dust-to-gas ratios
\fesc\ can be quite low, but their overall absorption remains weak 
($EW \sim -5$ to 0 \AA). These models, e.g., occupy the lower right
corner of Fig.\ \ref{fig_ewabs_fesc}.
Overall, models with a large EW in absorption correspond to
high column densities (cf.\ Fig.\ \ref{fig_ewabs_nh}). As the \lya\
escape fraction mostly reflects the transmission of photons close to
line center, \fesc\ decreases more rapidly than the equivalent width
of the absorption line increases (in absolute terms). This explains the
near vertical behaviour of the sequences of constant \nhi\ shown 
in Fig.\ \ref{fig_ewabs_fesc}.

Since in reality, the intrinsic UV spectrum of star-forming galaxies
is in general composed of a UV continuum plus \lya\ emission
\citep[see e.g.][ for synthetic spectra in this region]{schaerer08},
Fig.\ \ref{fig_ewabs_fesc} schematically shows how the two components
(line and continuum) are affected by radiation transfer effects.
Approximately, the resulting emergent line is then the superposition
of the absorption carved from the continuum plus the remainder 
of the transmitted \lya\ emission line (with a positive equivalent width).
To predict the detailed shape \lya\ profile, showing a complex diversity
as e.g.\ shown by \citet{verhamme06,schaerer08,dessauges09,vanzella10}, the full
results from our radiation transfer simulations need to be used.
We now briefly illustrate some profiles and variations with the model parameters.

\begin{figure*}[tb]
\centering
\includegraphics[width=8.8cm]{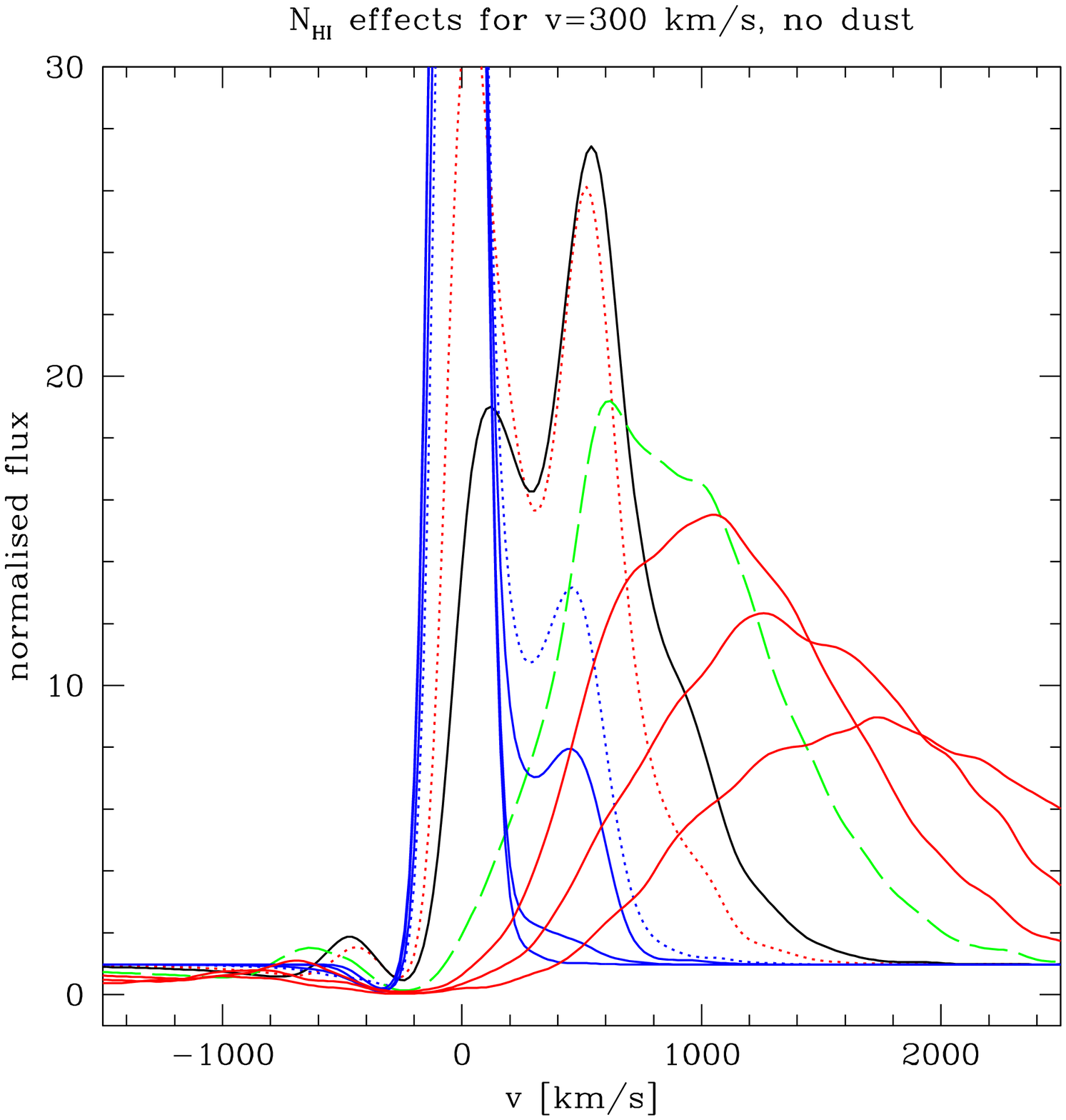}
\includegraphics[width=8.8cm]{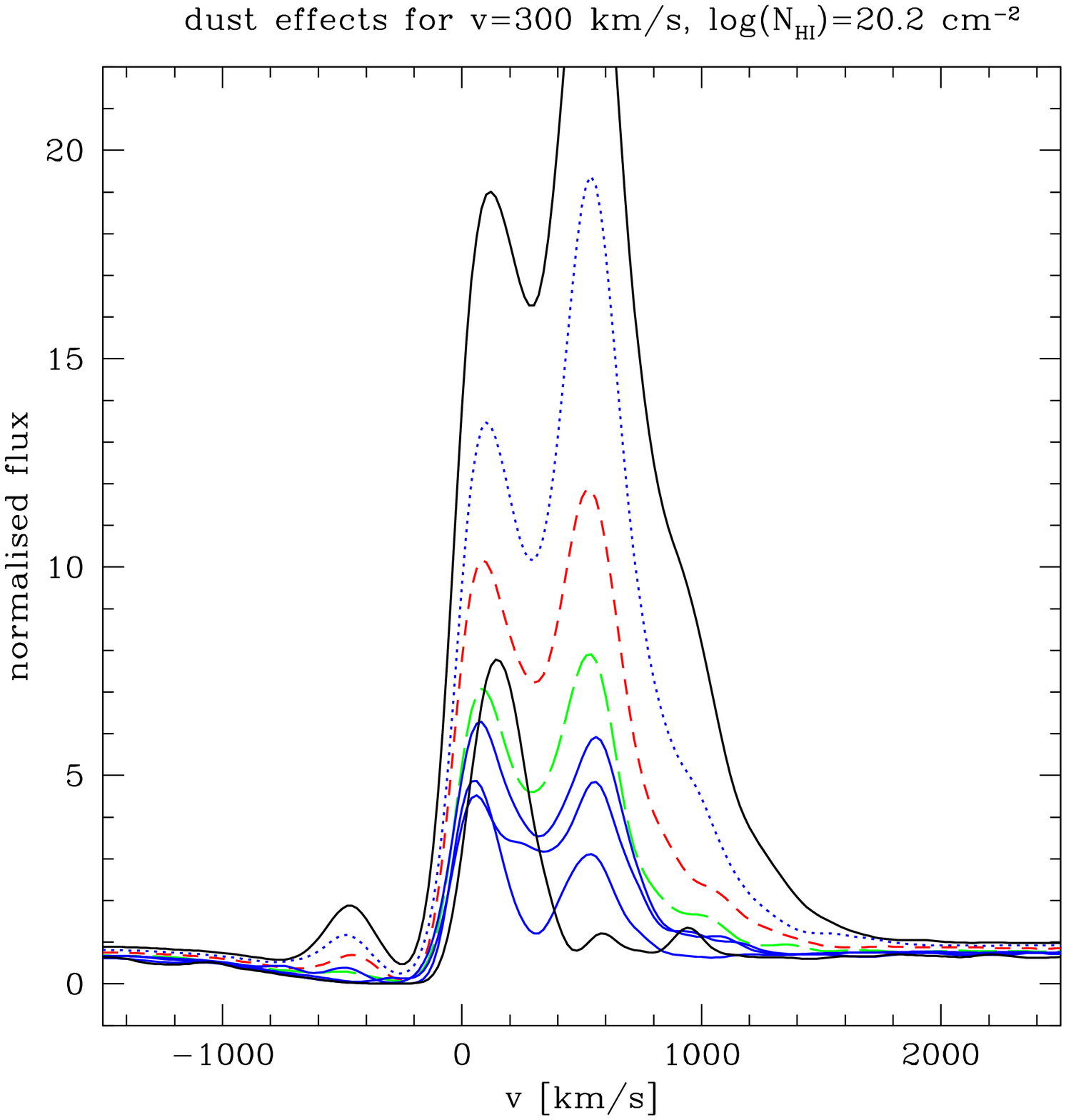}
\caption{Predicted normalised \lya\ line profiles for an expanding shell with
$\vexp = 300$ \kms, $b=40$ \kms, and with varying \hi\ column
densities (left panel), and varying dust content (right panel).
All line profiles are normalised around $\pm$ 5900 \kms.
{\bf Left:} Shown are models with $\log \nh=$ 19.3 (blue dotted), 19.9 (red dotted), 20.2 (black), 20.8 (green dashed)
\cm2\ comparable to those in Fig.\ 16 of \citet{verhamme06}, plus
models with lower column densities ($\log \nh=$ 16., 18., 19. \cm2, solid blue), and with
higher lower column densities ($\log \nh=$ 21.1, 21.4, 21.7 \cm2, solid red).
{\bf Right:} Shown are models with $\taua=0.$ (black, top), 0.1 (blue dotted), 0.5 (red dashed),
1. (green dashed) comparable to those in Fig.\ 16 of \citet{verhamme06}, plus
models with higher optical depths ($\taua=1.5$, 2., 3. (solid blue),and 4. (solid black)).
}
\label{fig_16}
\end{figure*}

\begin{figure}[tb]
\centering
\includegraphics[width=8.8cm]{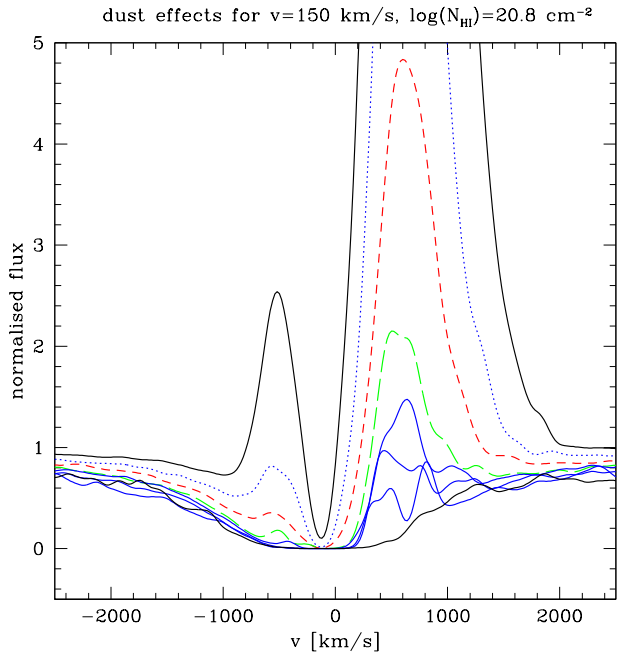}
\caption{Predicted normalised (at $\pm$ 5900 \kms) \lya\ line profiles for an expanding shell with
$\vexp = 150$ \kms, $b=40$ \kms, $\log \nh=20.8$ \cm2, and varying dust content
described by $\taua=0.$, 0.2, 0.5, 1., 1.5, 2., 3. and 4 (from top to bottom).
Note how the increasing dust content progressively reduces the line to ``carve'' out 
a damped (Voigt-like) profile for the highest dust contents.}
\label{fig_dust}
\end{figure}

\begin{figure}[tb]
\centering
\includegraphics[width=8.8cm]{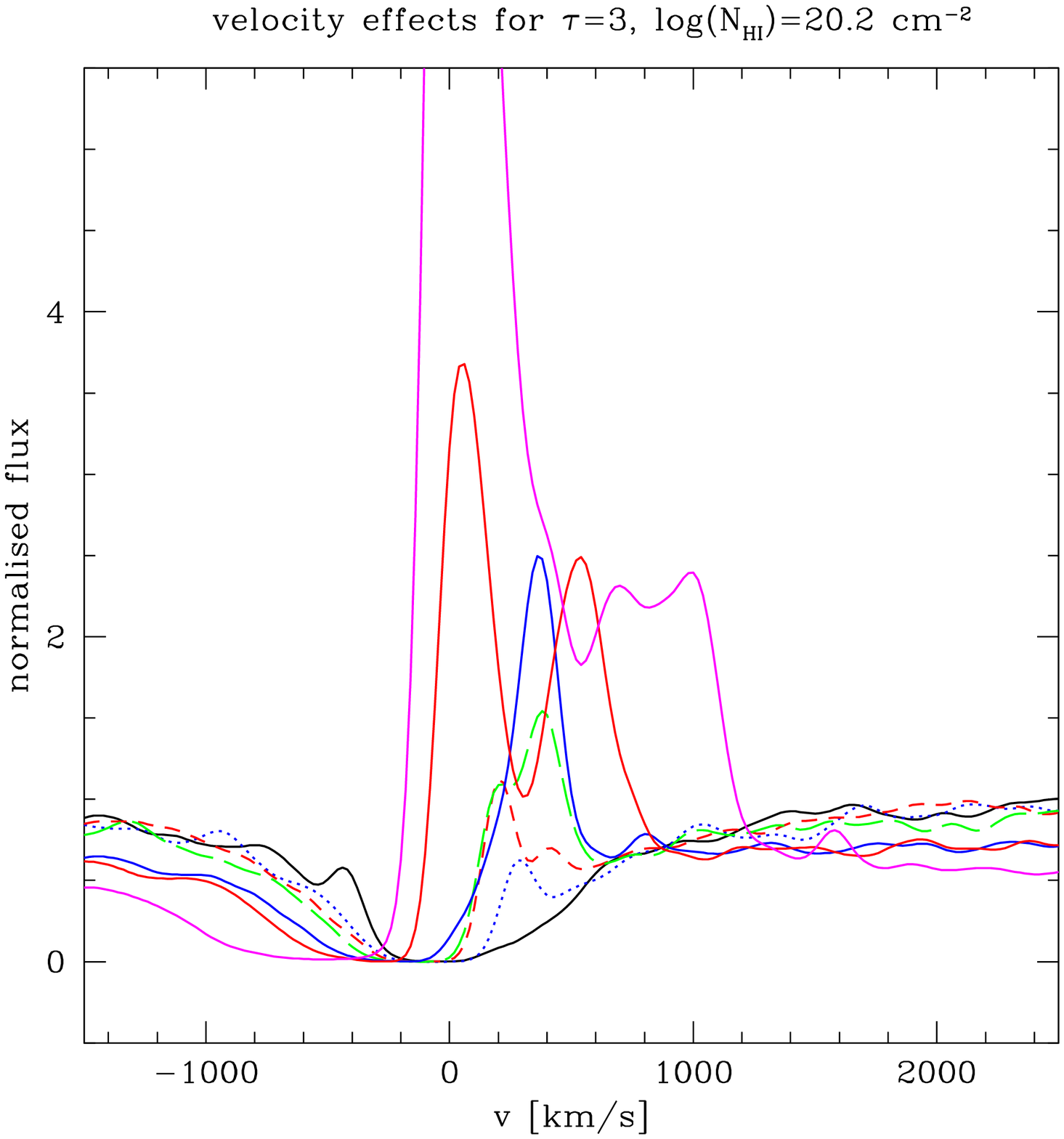}
\caption{Predicted normalised (at $\pm$ 5900 \kms)
\lya\ line profiles for an expanding shell with
$\taua = 3$ \kms, $b=40$ \kms, $\log \nh=20.2$ \cm2, and varying shell velocity
$\vexp = 0$ (black), 20 (blue dotted), 50 (red dashed), 100 (green dashed), 200 (blue solid), 
300 (red solid), 600 (magenta) \kms.}
\label{fig_highdust_v}
\end{figure}

\section{A library of theoretical \lya\ line profiles}
\label{s_profile}
As already mentioned, the results from our radiation transfer models
can be used to compute {\em a posteriori} the predicted spectrum around
\lya\ for any arbitrary input spectrum. 
To illustrate some \lya\ line profiles predicted by our extensive model grid,
we show few selected results computed for a continuum plus \lya\ emission,
assuming FWHM$=100$ \kms\ and \ewlya $=60$ \AA\ (except mentioned otherwise),
and plotted for a spectral resolution of 150 \kms.

In Fig.\ \ref{fig_16} we show simulations for a rapidly expanding shell
($\vexp = 300$ \kms, assuming $b=40$ \kms) with varying \hi\ column
densities (left panel), and varying dust content (right panel).
Both panels are similar to Figs.\ 16 and 17 of \citet{verhamme06}, but we
here show the predictions for a wider parameter space.
Qualitatively the behaviour is as discussed and explained by \citet{verhamme06}.
In particular, for models with different column densities, it is seen that 
the position of the most prominent peak of the profile shifts to 
$\ga (1.5-2) \times \vexp$ for $\log \nh \ga 10^{20}$ \cm2\ (left panel).
For the highest column densities the profile becomes very flat and broad,
extending to very high velocities ($> 1500$ \kms).
The right panel illustrates the effect of dust (absorption) optical depths
up to $\taua=4$. While to first order the profile shape remains similar
(when renormalised) but somewhat ``sharpened'' for relatively low optical depths 
\citep[$\taua \la 1$, cf.][]{verhamme06,verhamme08,laursen09a},
the overall shape of the profile can be more significantly altered
at higher optical depths. Here, e.g., the main peak gets nearly destroyed
for $\taua=4.$.

Fig.\ \ref{fig_dust} shows the line profile from an outflow with 
$\vexp = 150$ \kms, a relatively high column density ($\log \nh=20.8$ \cm2),
and for varying dust content ($\taua=0$ to 4.).
Note how the increasing dust content progressively reduces the line to ``carve'' 
out a damped (Voigt-like) profile for the highest dust contents.
Such damped profiles with some remaining \lya\ emission in the wings
have e.g.\ been observed for relatively dusty LBGs such as cB58 and the
8 o'clock arc, which have successfully been modeled with simulations
from this grid \citep[cf.][]{schaerer08,dessauges09}.

As a final illustration of the large parameter space covered by our
models, we show in Fig.\ \ref{fig_highdust_v} how the \lya\ profile varies with 
shell velocity in a very dusty, large column density environment.
As expected, an emission line emerges out of the damped profile seen for a static shell 
(black) when the shell has a net radial flow velocity, and quite rapidly
the profile becomes resembles a P-Cygni profile with blue shifted absorption
and redshifted emission, with the complex detailed shape whose origin
has been discussed in \citet{verhamme06}.

Again, note that our results also apply for symmetry reasons to models with 
spherical infall instead of outflows. In this case the predicted line profiles
are simply inverted in velocity space.

To make efficient use of this large spectral library we (MH) have also developed
an automatic fitting tool, which has already been applied in \citet{dessauges09,vanzella10,lidman11}.
Other applications will be presented elsewhere.

\section{Conclusions}
\label{s_conclude}

We have improved our 3D \lya\ and UV continuum radiation transfer code
\mclya\ \citep{verhamme06} and parallelised it for efficient
use. Using this code we have computed a large grid of radiation
transfer models for homogeneous spherically expanding 
(and by symmetry also infalling) shells
containing \hi\ and dust surrounding a central source. With 6240
simulations, the grid covers a large parameter space with radial
velocities from 0 to 700 \kms, neutral hydrogen column
densities \nh\ from $10^{16}$ to $10^{21.7}$ \cm2, Doppler parameters
$b=10$ to 160 \kms, and dust optical depths $\taua=0.$ to 4
(corresponding approximately to \ebv $\sim$ 0.4).

From the simulations we determine the predicted \lya\ escape fraction, 
the UV continuum attenuation, and we illustrate some of the \lya\ line profiles
which are predicted by these models. The model predictions provide a useful
basis for the interpretation of \lya\ observations (line fluxes, 
escape fractions, equivalent widths, detailed line profiles and others) from
star-forming galaxies, including Lyman alpha emitters (LAEs), Lyman break galaxies
(LBGs) and others. The models have already been successfully applied 
and confronted to a variety of observations
\citep[see e.g.][]{atek09b,hayes10b,hayes10c,vanzella10,dessauges09,garel11} 
and are now made available publicly.

Our models should also serve as an important base for comparison for future
models including other refinements (e.g.\ other geometries and velocity fields, 
non-homogeneous media) and further improvements.

\begin{acknowledgements}
Simulations were done on the {\tt regor} PC cluster at the Geneva
Observatory co-funded by grants to Georges Meynet, Daniel Pfenniger, and DS,
and on the Bull platine at the CEA.
We would like to thank the granted access to the HPC resources of CINES and CCRT under 
the allocations 2009-SAP2191 and 2010-GEN2192 made by GENCI.
The work of MH, DS, and AV was supported by the Swiss National Science Foundation.

\end{acknowledgements}

\bibliographystyle{aa}
\bibliography{references}

\end{document}